# Hybrid Cerenkov-scintillation detector validation using Monte Carlo simulations

Emilie Jean[1,2,3], Simon Lambert-Girard[4], François Therriault-Proulx[4] and Luc Beaulieu[1,2]

[1] Département de physique, de génie physique et d'optique et Centre de recherche sur le cancer, Université Laval, Quebec, QC, Canada
[2] Département de radio-oncologie et Axe Oncologie du CRCHU de Québec, CHU de Québec - Université Laval, Quebec, QC, Canada
[3] Département de radio-oncologie du CIUSSS-MCQ, CHAUR de Trois-Rivières, Trois-Rivières, QC, Canada
[4] Medscint inc. Quebec, QC, Canada

**Abstract**

**Objective:** This study aimed at investigating through Monte Carlo simulations the limitations of a novel hybrid Cerenkov-scintillation detector and the associated method for irradiation angle measurements.
**Approach:** Using Monte Carlo simulations, previous experimental irradiations of the hybrid detector with a linear accelerator were replicated to evaluate its general performances and limitations. Cerenkov angular calibration curves and irradiation angle measurements were then compared. Furthermore, the impact of the Cerenkov light energy dependency on the detector accuracy was investigated using the energy spectra of electrons travelling through the detector.
**Main results:** Monte Carlo simulations were found to be in good agreement with experimental values. The irradiation angle absolute mean error was found to be lesser than what was obtained experimentally, with a maximum value of $1.12°$ for the 9 MeV beam. A 0.4% increase of the ratio of electrons having an energy below 1 MeV to the total electrons was found to impact the Cerenkov light intensity collected as a function of the incident angle. The effect of the Cerenkov intensity variation on the measured angle was determined to vary according to the slope of the angular calibration curve. While the contribution of scattered electrons with a lower energy affects the detector accuracy, greatest discrepancies result from the limitations of the calculation method and the calibration curve itself.
**Significance:** A precise knowledge of the limitations of the hybrid detector and the irradiation angle calculation method is crucial for a clinical implementation. Moreover, the simulations performed in this study also corroborate hypotheses made regarding the relations between multiple Cerenkov dependencies and observations from the experimental measurements.

## 1. Introduction

Modern radiotherapy treatments employ complex technological strategies such as intensity modulated radiotherapy (IMRT, VMAT), stereotactic radiosurgery (SRS, SBRT)(Meyer 2011) and magnetic resonance image-guided radiotherapy (MRgRT) (Liney and Heide 2019). While these techniques have shown to provide better results by maximizing the dose to the tumour and sparing surrounding tissues, they also increase the complexity of dose measurements and highlight the limitations of existing dosimeters (Ezzell *et al* 2003, Benedict *et al* 2010, Solberg *et al* 2012, Jelen and Begg 2019, O'Brien *et al* 2018). Consequently, these techniques have motivated research to develop new, more efficient detectors.

Previous characterization of a novel Cerenkov-scintillation detector (Jean *et al* 2022) has demonstrated it could be a convenient tool for complex dosimetry conditions. The detector is designed to simultaneously measure the deposited dose and the mean incident angle of electrons striking the detector, which can be linked to the irradiation angle of external photon and electron beams, the source position for brachytherapy treatment or the effect of a magnetic field when using a MR-LINAC. While also measuring the deposited dose, the hybrid Cerenkov-scintillation detector provides useful additional information that other conventional dosimeters are unfit to measure. First results offer promising perspective for future applications in external beam radiotherapy, especially for stereotactic radiosurgery (SRS) and magnetic resonance image-guided radiotherapy (MRgRT).

Therefore, the main objective of this technical note was to investigate, via Monte Carlo simulations, experimental measurements performed to establish a proof-of-concept of the hybrid Cerenkov-scintillation detector. The precise knowledge of the efficiency and general performance of the detector is crucial for its clinical use as a tool for irradiation angle measurements. In turn, the knowledge of the spatial and angular distributions, and dynamics of Cerenkov photons emitted within an optical fibre is of great interest for the understanding of the detector limitations. In order to determine the general performance and limitations of the hybrid detector, results from the experimental investigations from Jean *et al*. (Jean *et al* 2022) using a Varian Clinac ix (Varian Medical, Palo Alto, USA) were compared with MC computations, which have been used for decades in the field of radiotherapy dosimetry (Andreo 2018). This served the purpose of corroborating hypotheses made regarding the relations between multiple Cerenkov dependencies and observations from the experimental measurements.

## 2. Material and methods

### 2.1 Hybrid Cerenkov-scintillation dosimeter

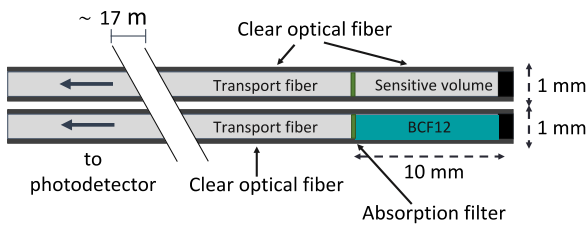

*Figure 1: Design schematic of the hybrid detector showing the Cerenkov and the scintillation probe assemblies (Jean et al 2022).*

The hybrid Cerenkov-scintillation detector developed in the previous study (Jean *et al* 2022) was designed as two distinct probes, that is a single-point scintillation dosimeter and a Cerenkov detector for angular measurements. The latter was built using a 10 mm long optical fibre with a PMMA-based core and fluorinated polymer-based cladding sensitive volume (1 mm diameter ESKA GH-4001, Mitsubitshi Chemical Co., Tokyo, Japan) separated from the transport fibre by an absorptive filter. This allowed to modify the Cerenkov spectrum produced in the sensitive volume and therefore, an algorithm could properly eliminate the contamination signal from the transport fibre (ESKA GH-4001). The former was composed of a 10 mm long BCF12 scintillating fibre (Saint-Gobain, Hiram, USA) and was used simultaneously to provide a real-time dose measurement and account for the Cerenkov electron energy dependency.

### 2.2 Monte Carlo

The *TOol for PArticle Simulation* application (TOPAS, version 3.7.0) developed by Perl *et al*. (Perl *et al* 2012, Faddegon *et al* 2020) was used to conduct Monte Carlo simulations. It wraps and extends the *Geant4* code (version 10.06.p03) (Agostinelli *et al* 2003, Allison *et al* 2006) and provides a user-friendly MC toolkit with no need to write C++ code.

To avoid the tracking of Cerenkov photons produced in the entire geometry, thus increasing the computing time, simulations were divided in two different processes. Ionizing radiation simulations were initially conducted to score the electron energy spectrum and the deposited dose inside the sensitive volume of the Cerenkov probe while also saving the phase space of all particles entering the same volume. Then, optical simulations were performed to score Cerenkov optical photons resulting from the charged particles travelling through the sensitive volume. All simulations were carried out on a MacBook Air laptop equipped with a 2.0 GHz i7 core. Each calculation took, on average, 6 hours to complete for the radiative part while each of them took 25 minutes for the optical part.

#### 2.2.1 Ionizing radiation simulations

To replicate the measurement conditions used for the detector irradiations, Varian Clinac ix (Varian Medical, Palo Alto, USA) phase spaces from PRIMO (Rodriguez *et al* 2013, Sarin *et al* 2020) were used as particle sources in TOPAS. The latter were scored below the applicator for electrons beams of 6, 9, 12, 16 and 20 MeV. Each of them contained $1 \times 10^8$ events, which were all transported to achieve ionizing radiation simulations. For all energies tested, phase spaces of $10 \times 10$ cm$^2$ field size were used for angular calibration curve simulations while $20 \times 20$ cm$^2$ were used for electron beams irradiation angle measurements. Phase space coordinates in TOPAS were set according to the scoring distance from the source to ensure the isocentre was located at 100 cm from the original source.

The geometry components for the ionizing radiation simulations included both probes of the detector embedded in a solid cylinder-shaped water-equivalent phantom. The compositions and dimensions of all elements of the probes were defined according to the manufacturer specifications. A first phantom of 6 cm diameter was modelized for electron beams of 12 to 20 MeV. A smaller cylinder of 3 cm diameter was used for lower energies. Sensitive volumes were centred at the isocentre and pointing toward the X2 jaw for angular calibration curve measurements. They were then shifted 8.5 cm along the lateral axis for uneven dose measurements as a function of the irradiation angle as illustrated in figure 2. Gantry angles ranging between 0˚ and 180˚ by increments of 5˚ were simulated for all energies.

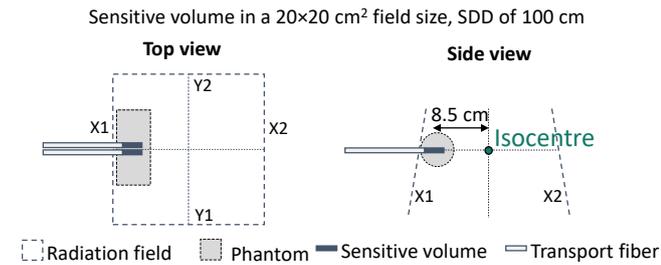

*Figure 2: Schematic top and side views of the set-up used for irradiation angle measurements showing the coordinate system. The source-detector distance (SDD) is calculated with the gantry positioned at 0˚.*

The g4em-standard_opt3 (in TOPAS nomenclature) physics list constructor was included in the simulations. The electron energy cut-off was set at 50 keV as the fibre diameter was larger than the range of such electrons in PMMA. No variance reduction technique was used. The absorbed dose in the sensitive volume was calculated for all irradiation conditions. Energy spectra of charged particles travelling through the sensitive volume were also obtained to evaluate the effect of a change in the energy distribution on the irradiation angle accuracy. The spectra were scored for energies ranging from 178 keV to 21.178 MeV using a binning of 100 keV. For all quantities, statistical uncertainties estimated by the history-by-history method were obtained using the built-in algorithm from TOPAS (Knuth 1997). Standard deviation (1σ) associated with the distribution of the quantity of interest was then multiply by the square root of the total number of histories to obtain the standard deviation of the summed quantity. A phase space of all particles above the Cerenkov energy threshold in PMMA (178 keV) entering the volume was also saved to further conduct optical simulations. The scoring surface was set as any surface of the Cerenkov sensitive volume and included particles strictly going in.

*2.2.2 Optical simulations*

The geometry used for optical simulations consisted of the Cerenkov probe placed in a 30x30x30 cm$^3$ world of air with the phase spaces from the ionizing radiation simulations superimposed over the sensitive volume. The GEANT4 optical processes (G4OpticalPhysics) were included to the physics list class to enable optical photons production, tracking and scoring. Cerenkov photons dynamic (G4Cerenkov class) is characterized by the charged particle properties but also optical properties of the media they are travelling through. Thus, the fibre core and cladding respective compositions, absorption lengths and refractive indexes (Zhang *et al* 2020) for wavelengths ranging from 350 nm to 700 nm were defined. Refractive index of air for the same wavelength range was set to 1 (Mathar 2007) to enable its optical behaviour, thus allowing total intern reflexion within the fibre. This prevented photons from escaping the cladding outside surfaces and the exposed core at the far end of the sensitive volume. All surfaces of the probe were set as smooth and perfectly polished. A surface having a dielectric-metal optical behaviour was also placed at the exit of the 17 cm long transport fibre to act as a photodetector. As the GEANT4 optical processes generate Cerenkov photons only within the wavelength range where the user has provided an index of refraction, all optical photons originating from the sensitive volume and reaching the photodetector were scored. Consequently, optical photons resulting from the contamination signal produced in the transport fibre were eliminated.

*2.3 Comparison with experimental measurements*

Analysis of the collected data was done by combining results from the ionizing and optical simulations. Angular calibration curves for all energies were obtained using the dose scored in the Cerenkov probe sensitive volume for a 10x10 cm$^2$ field size and corresponding Cerenkov light intensity as a function of the irradiation angle. Matching doses and Cerenkov light intensities for uneven doses as a function of the gantry angle where then used to solve the calibration function and calculate the irradiation angle using the 3-steps process described by Jean *et al.* (Jean *et al* 2022). Results of the experimental measurements from the same study were finally compared with those obtained from Monte Carlo calculations.

**3. Results and discussion**

Using the built-in statistical uncertainty estimator, the deposited dose in the sensitive volume for a given irradiation condition was found to have a maximal deviation of 0.6% for all energies tested. The absorbed dose as a function of the gantry angle was also found to be constant within 0.7% using various beam energies. No trend was observed as a function of the irradiation angle.

Example of the sensitive volume Cerenkov signal from MC calculations as a function of the irradiation angle for fixed dose are illustrated in figure 3.a and figure 3.b for two different energies. Corresponding experimental values are also shown for comparison. For all energies, a maximal difference of 16.4% of the Cerenkov relative intensities can be seen between experimental and MC. The greatest discrepancies occur for gantry angles where the beam is perpendicular to the fibre axis or pointing toward the end of the probe. As all surfaces are defined as perfectly polished for the MC calculations, the reflexions from the optical boundaries located at the tip of the probe are higher compared to the experimental probe. Comparison with ground surfaces as per GEANT4 nomenclature (G4OpticalSurfaceFinish) has shown to reduce up to 6.7% the ratio of collected light at normal incidence with respect to the fibre axis. Other discrepancies are caused by the simplified model, including the reduced transport fibre length and absence of refractive index discontinuities between the detector components. However, trend of the Cerenkov calibration curves as a function of the beam energy from MC simulations is similar to that from experimental values.

*Table 1 : Angle of irradiation absolute mean and max errors for experimental data (Jean et al 2022) and Monte Carlo calculations using 9, 12, 16 and 20 MeV electron beams.*

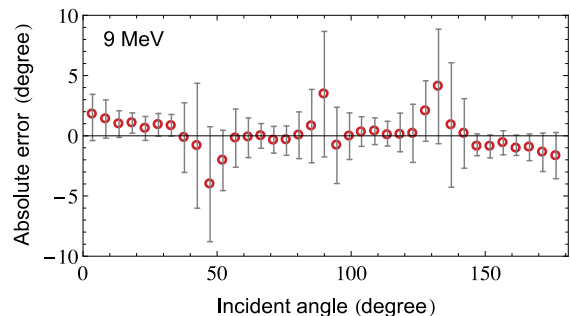

|  |  | 9 MeV | 12 MeV | 16 MeV | 20 MeV |
|---|---|---|---|---|---|
| Experimental | Mean | 1.97° | 1.66° | 1.45° | 0.95° |
|  | Max | 6.52° | 4.30° | 3.29° | 2.85° |
| Monte Carlo | Mean | 1.12° | 0.95 | 0.92° | 0.75° |
|  | Max | 4.10° | 3.08° | 2.97° | 2.21° |

Comparison of the irradiation angle absolute mean errors for MC and experimental measurements are listed in table 1 for electron beams of 9 to 20 MeV. The 6 MeV beam could not be used for incident angle measurements due to its poor angular dependency, providing a curve with regions having a slope of zero. MC simulations result in better accuracy, with a maximal mean absolute error of 1.12°. Similarly, higher beam energies also yield better results. Experimental values are in close agreement with those obtained from the Monte Carlo calculations. The largest errors are found at parallel incidence with respect to the fibre axis (i.e., gantry at 90°) and around the peak intensity angles as illustrated in figure 4 for the 9 MeV electron beam, which is the energy with the highest discrepancies. Regions of the calibration curve where the slope is lower, or its sign changes tend to provide larger errors as a small difference in the measured intensity will greatly modify the calculated angle. The curve, being interpolated between measurements, could also be less precise for those regions. Conversely, regions where the Cerenkov intensity normalized to the dose differ the most from the calibration curves, as illustrated in figure 5 for the 9 MeV beam, do not yield major errors on the irradiation angle calculation.

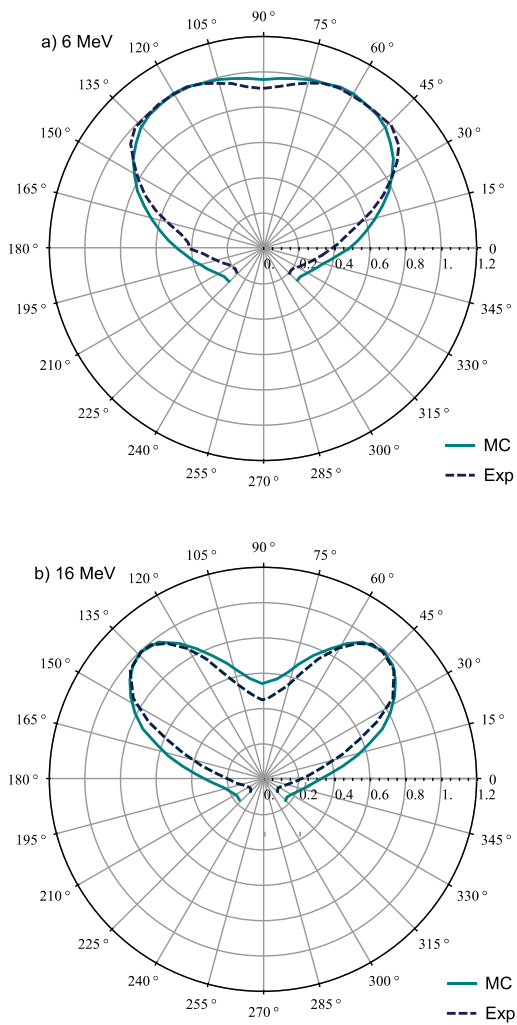

*Figure 3 : Signal emitted by the Cerenkov detector sensitive volume as a function of the gantry angle for fixed dose using (a) 6 and (b) 16 electron beams. Monte Carlo calculations (full line) and experimental measurements (dashed line) are normalized to their respective maxima. The detector is pointing toward the gantry at 90°.*

*Figure 4: Angle of incidence absolute error calculated with the MC simulations using a dose varying as a function of the irradiation angle for a 9 MeV electron beam. Error bars were calculated using uncertainties on the Cerenkov intensities.*

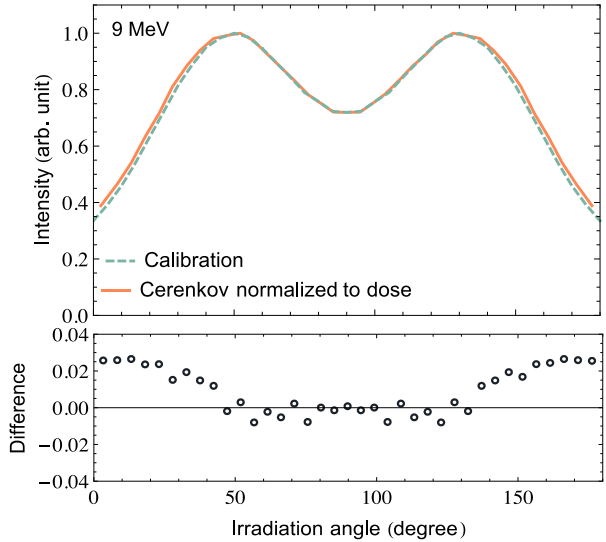

*Figure 5: Angular calibration curve and Cerenkov intensity normalized to dose for irradiation angle measurements obtained with a 9 MeV beam from MC simulations.*

Energy fluence spectra are further compared between calibration curve and uneven dose measurements. The ratio of electrons having an energy lower than 1 MeV to the total electrons traveling through the sensitive volume are shown in figure 6. The threshold is defined knowing that energy has a greater impact on the Cerenkov emission angle and yield for electrons of approximately 1.0 MeV or less according to Frank-Tamm and emission angle equations (Jelley 1958). As the particle energy becomes less relativistic and approaches the threshold energy, the phase velocity decreases significantly, inducing fewer Cerenkov photons with lower angle between the fibre axis and the particle path (Law *et al* 2006). Energy spectrum differences are found to increase with the off-axis distance, that is for gantry angles approaching 0°. The phantom being partially in the penumbra region due to its position with respect to the isocentre, a lack of scattered electrons can be seen for those angles. This is also where the measured Cerenkov intensity differs from the calibration curve (see figure 5). While a 0.4 % variation of the ratio of electrons of 1 MeV or less was found to have an impact of approximately 2% on the collected intensity, a change in the spectrum shape above 1 MeV did not impinge the results as long as the ratio to the total is the same.

## 4. Conclusion

In ideal light transport conditions, Monte Carlo simulations show the irradiation angle can be measured with a maximum absolute mean error of 1.12°, regardless of the energy employed. The differences between Monte Carlo and experimental results are caused by the simplified model of the detector in MC calculations as well as the uncertainties of the experimental results. However, the good agreement between both validates the hypotheses made regarding the proportionality of Cerenkov light and absorbed dose but also the influence of electron energy spectra on the collected intensity. While an increase as low as 0.4% of the ratio of electrons having an energy below 1 MeV has an impact on the collected intensity and the measured irradiation angle, most significant discrepancies arise from the interpolation of the calibration curve and the calculation method. Consequently, improvement of the latter could enhance the detector precision and reduce its limitations.

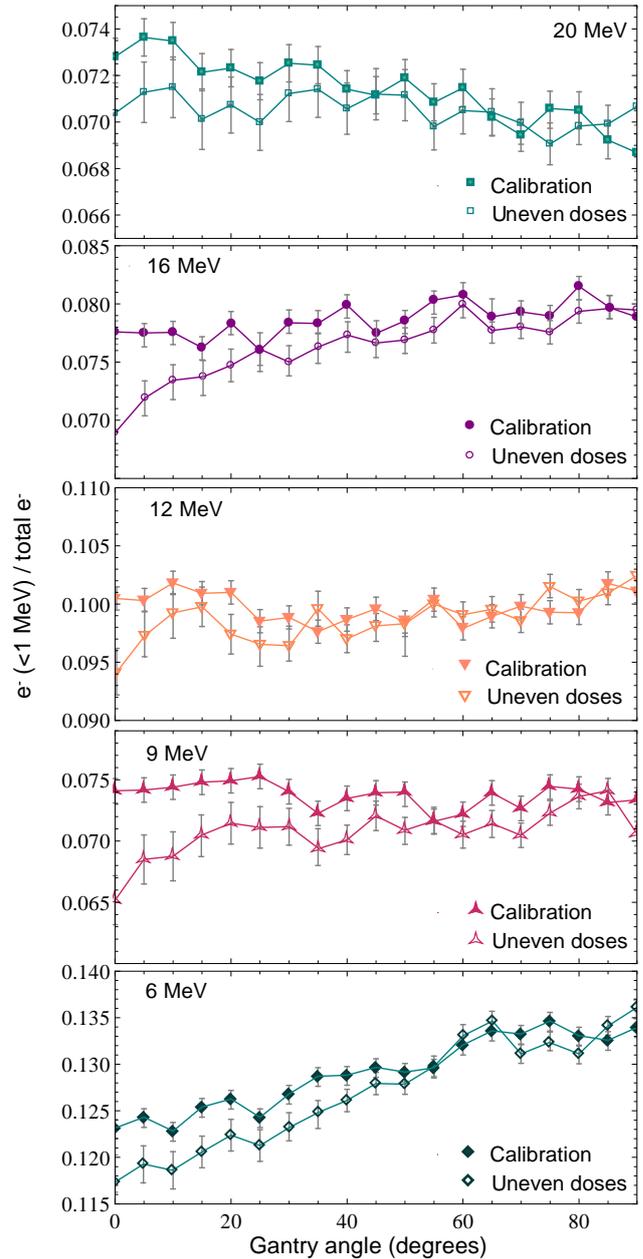

*Figure 6: Ratio of electrons having an energy lower than 1 MeV to the total electrons travelling through the Cerenkov probe sensitive volume obtained from angular calibration curve and uneven dose simulations for 6, 9, 12, 16 and 20 MeV electron beams.*

## Acknowledgements

This work was supported by the *Natural Sciences and Engineering Research Council of Canada* (NSERC) Discovery grants RGPIN 05038-2019, a MEI PSO2D Maturation grant (MEI_v-réf 53026 n-réf 12-453PSO2D) and the Fellowships Program of the *Ministère de la Santé et des Services Sociaux du Québec* (MSSS). François Therriault-Proulx and Simon Lambert-Girard are Co-founder at Medscint inc., a company developing scintillation dosimetry systems. This work was not financially supported by Medscint.